\documentclass[aps,pre,twocolumn,showpacs,floatfix]{revtex4}

\usepackage[utf8]{inputenc}
\usepackage{amsmath}
\usepackage{graphicx}
\usepackage{color}
\usepackage{bm}
\usepackage{amssymb}

\begin{document}

\bibliographystyle{apsrev}

\title{Radiative ablation with two ionizing fronts \\ when opacity displays a sharp absorption edge}
\author{\surname{Olivier} Poujade}
\affiliation{CEA, DAM, DIF, F-91297 Arpajon, France}
\email[Correspondance should be adressed to: ]{olivier.poujade@cea.fr}
\author{\surname{Max} Bonnefille}
\affiliation{CEA, DAM, DIF, F-91297 Arpajon, France}
\author{\surname{Marc} Vandenboomgaerde}
\affiliation{CEA, DAM, DIF, F-91297 Arpajon, France}

\date{\today}

\begin{abstract}

The interaction of a strong flux of photons with matter through an ionizing-front (I-front) is an ubiquitous phenomenon in the context of astrophysics and inertial confinement fusion (ICF) where intense sources of radiation put matter into motion. When the opacity of the irradiated material varies continuously in the radiation spectral domain, only one single I-front is formed. In contrast, as numerical simulations tend to show, when  the opacity of the irradiated material presents a sharp edge in the radiation spectral domain, a second  I-front (an edge-front) can form. A full description of the mechanism behind the formation of this edge-front is presented in this article. It allows to understand extra shocks (edge-shocks), displayed by ICF simulations, that might affect the robustness of the design of fusion capsules in actual experiments. Moreover,  it may have consequences in various domains of astrophysics where ablative flows occur.

\end{abstract}

 \maketitle

\section{Observation : two shocks when only one is expected}

The rocket effect produced by the ablation of matter submitted to an intense radiation flux was first introduced in the fifties by Oort and Spitzer \cite{oort} in the astrophysical context to explain the radiation-driven formation of O-star. At the same time, Kahn \cite{kahn, mia} was the first to theorize and classify the various possible ionization-fronts (I-fronts in \cite{kahn} and often referred to as radiation-front in the ICF community), responsible for the phenomenon of ablation, with respect to the intensity of the ionizing radiation. In a nutshell, when the intensity is small enough, an I-front propagates subsonically upstream in the un-ionized material (in Kahn's terminology, it is a weak D-front). As the intensity increases, the I-front catches up with the rarefaction and overtakes it (it becomes D-critical). An additional increase of the photon flux makes the pressure at the I-front (ablation pressure) strong enough to trigger {\it one single} shock that propagates upstream in the un-ionized material. The front is then called a D-front and the shock is said to be ablatively-driven (it is the physics behind the creation of such a shock that is our main concern here). If the intensity is increased yet further, the I-front overtakes the shock (it is R-critical) and can even propagate supersonically in the un-ionized material (it is then called an R-front). \par
The goal of this paper is to show that, under certain conditions, two I-fronts can be generated by an intense radiation flux instead of one. This feature might have consequences in various domains of astrophysics where ablatively-driven flows play a significant role, such as star formation (as already mentioned), acceleration of interstellar clouds \cite{massload}, formation of Strömgren sphere in gaseous nebulae \cite{strom} and supernovae remnant \cite{sn}. However, it is in the domain of inertial confinement fusion (ICF) that this phenomenon has been highlighted. 

\par 

Ablatively-driven shocks play a role of paramount importance in ICF, for a promising way of achieving thermonuclear ignition, on the National Ignition Facility (NIF)  and the Laser MegaJoule (LMJ), relies on the indirect-radiative-drive of a small plastic capsule filled with a mix of deuterium and tritium (DT) \cite{nif}. In this context, a precisely shaped laser pulse is converted into x-rays in a high-Z material (gold) Hohlraum at the center of which the capsule is located. These x-rays ionize the spherical capsule's outer surface (plastic ablator) and a spherical D-front builds up in reaction. Each time the radiative temperature of the x-ray flux undergoes a sudden, discontinuous change of slope, the D-front generates a spherical shock converging to the center of the capsule and dragging down the shocked material along with it. 

\par 

In order to compress the capsule until the temperature and the areal density of the DT are high enough to trigger its ignition, a succession of three or four of these shocks is necessary. Their intensity and chronometry are adjusted \cite{keyhole} to minimize the entropy deposited in the cryogenic DT \cite{stiming}: the energy invested in the capsule must be devoted  to the actual useful compression (kinetic energy) and not to the detrimental heating of matter. 

\par

In numerical simulations of ignition capsule implosions with plastic ablators, it has been observed that a shock generated classically upstream of a D-front may come with a delayed partner that is generated in the ablated region (downstream of the D-front) at the very same time. It has been observed in 1D HYDRA simulations in \cite{nif1,nif2} and in our own simulations with FCI2  \cite{fci} but it has not been observed experimentally. 

\section{Motivation}

One benefit of simulations is the complete knowledge of density, pressure, temperature, velocity, etc. at each instant and everywhere in the whole simulated flow. It contrasts with the knowledge of the experimental flow, in general, which is confined to a reduced number of physical quantities over a more limited time and a smaller part of the flow. This is the reason why, sometimes, simulations reveal phenomenological issues that would be difficult if not impossible to observe experimentally. 

\par

The question that naturally arises concerning that delayed shock is why has it not been observed experimentally (in keyhole \cite{keyhole}) so far? The answer, once again, is suggested by simulations: the delayed shock could be hidden from the keyhole experiment because it collapses with a main sequence shock before being visible (The VISAR only records the velocity of the shock closest to the center of the ICF capsule). Of course, simulations have limitations of their own. The lack of precision of tabulated or modeled equations of state (eos) and opacities that feed the simulation code, to cite a few, could greatly impair the agreement with experiments. 

\par 

The subject of concern here is the fact that the delayed shock is very sensitive to these eos and opacities and much more sensitive than main sequence shock. Therefore, the simulated prediction of the behavior of this delayed shock is marred by greater uncertainties than main sequence shocks and can affect drastically the simulated shock timing. Fortunately, the actual shock timing of an ICF capsule is adjusted experimentally through the VISAR velocity profile recorded by a keyhole platform \cite{keyhole}. It is not set by means of simulation. In contrast, the post-shot modeling of ICF experiments relies heavily on simulations \cite{post1}. It is an inductive analysis that allows to go back from the effect (experimental observation of the VISAR velocity profile for instance) to the cause (radiative temperature ($T_r$) law as seen by the capsule) by adjusting the simulation to fit the experiment. But, this induction makes sense as long as a given effect cannot be produced by several different causes. Equations of state, opacities, NLTE model and many more data can be constrained through such type of indirect, inductive process : in \cite{nif1}, for instance, NLTE models are constrained in order to improve the agreement with the VISAR data, whereas in \cite{eos2}, eos for CH (LEOS 5370 and 5400) are constrained with the measured perturbation
growth factors for optical-depth modulations as a function of the
modulation mode number (Fig.5 in \cite{eos2}). 

\par

As will be proved later on in this paper, the problem with the delayed shock in question is that it is created in the ablated (underdense) plastic at densities ($\approx$ 0.05 g/cm$^3$) less than the standard value ($\approx$ 1.0 g/cm$^3$) whereas main sequence shocks are triggered in the compressed region of the ablation front where densities are above the standard value ($\approx$ few g/cm$^3$). Now, it is precisely in this underdense region that equations of state are not well characterized : experiments devoted to equations of state measure shocked states of matter which have densities above the standard value \cite{eos1} (to the authors knowledge there is no experiment constraining eos in the underdense regime). Two equations of state with different underdense behavior (but with the same normal-and-above-normal-density behavior) would not affect the main sequence shock but would trigger potentially different edge-shocks. It would then be possible to adjust the Tr law, as seen by the capsule, in two different ways (depending only upon the underdense behavior of the ablator) that would both be compatible with the same VISAR data. The conclusion of the postshot modeling \cite{post1} would then be unreliable. Therefore, a better characterization of underdense eos would further constrain the postshot modeling effort. 

\par

Here, we modestly want to clarify the mechanism behind the formation of these edge-shocks in order to understand their interplay with underdense eos. As will be proved later on in this article, this extra shock is the result of the formation of a second "I-front" downstream of the main ionization-front (ablation-front) owing to the singular shape of the opacity of carbon (which is the main contributor to the opacity of an undoped plastic ablator) with a K-edge around 0.3 keV. 

\begin{figure}[b] 
\includegraphics[width=8cm]{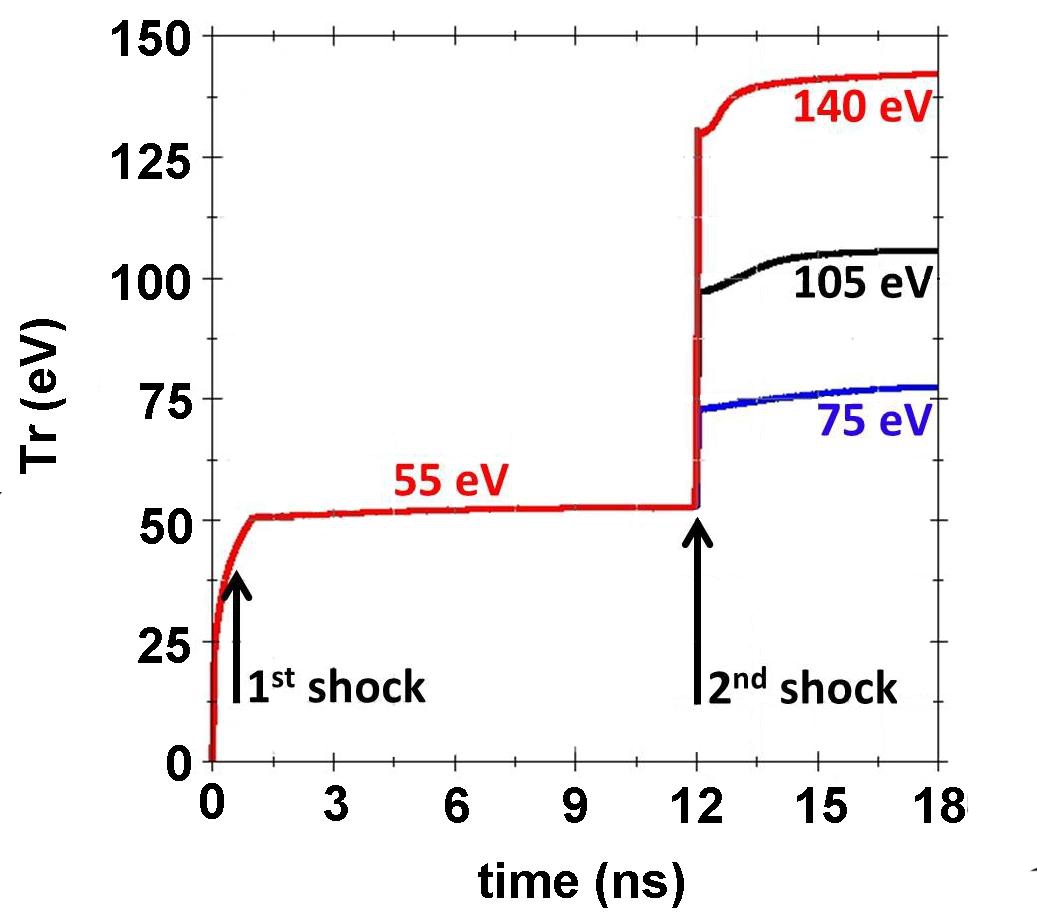}
\caption{(Color online) Three idealized $T_r$ laws of the ionizing radiation hitting the 300 $\mu$m ablator. The first rise is responsible for shock 1 and the second rise is responsible for shock 2 in Kahn's theory. Orders of magnitudes are typical of a NIF drive although the rises last longer and can be followed by small overshoots known as pickets that are irrelevant for the issue at stake here.}\label{fig:tr}
\end{figure}

\begin{figure}
\includegraphics[width=7.cm]{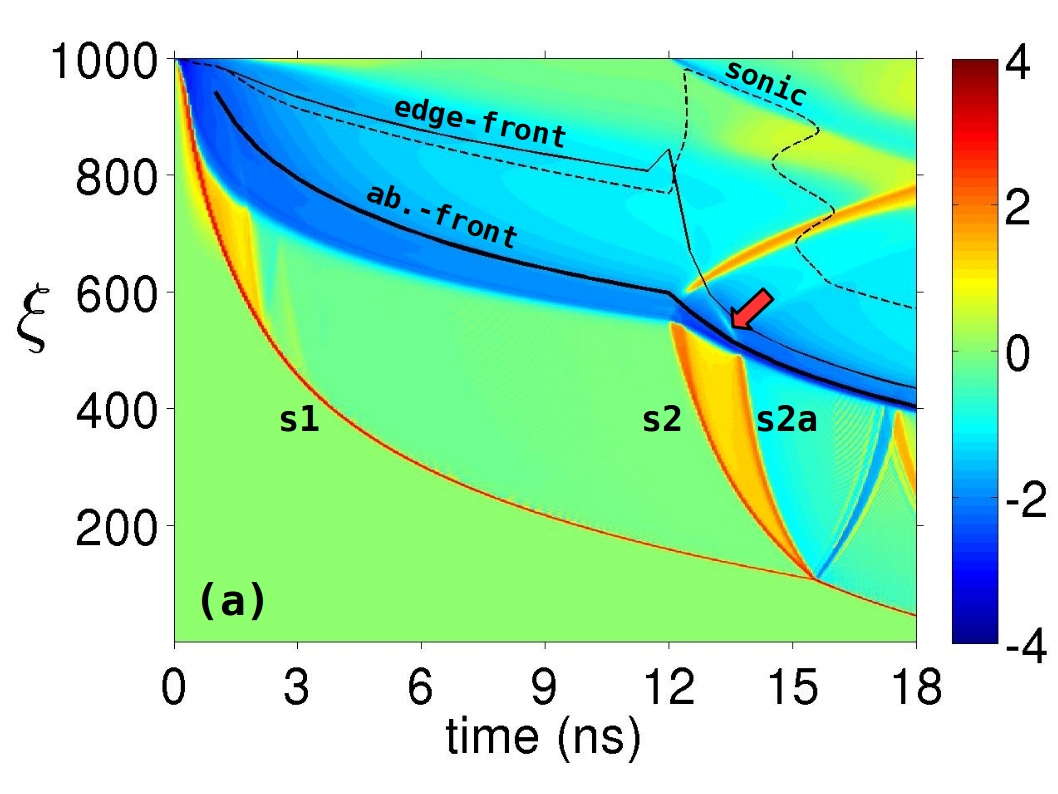}\\
\includegraphics[width=7.cm]{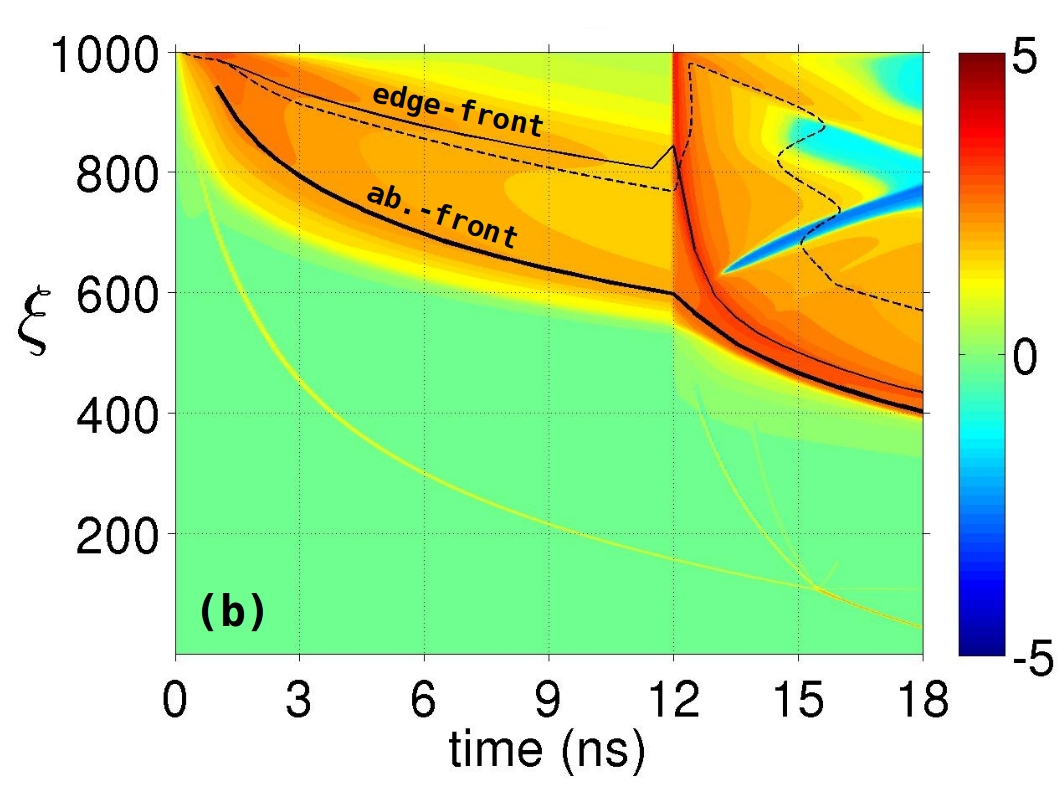}
\caption{(Color online) Color plots (when $T_r^\mathrm{max}$=105 eV) of $f\left[\mathrm{d}(\ln(\rho))/\mathrm{d}t\right]$ (up) and $f\left[T\mathrm{d}s/\mathrm{d}t\right]$ (down) with respect to time and lagrangian depth coordinate $\xi$. The function $f[x]=\mathrm{sgn}(x)\ln(1+|x|/x_0)$ allows to represent these physical quantities in log-scale with their sign. The black dashed line is the locus of sonic points with respect to the ablation-front. The thin black line is the penetration depth of photons with energy above 0.3 keV (energy of the K-shell of carbon) whereas the thick black line is the penetration depth of photons with energy below 0.3 keV (follows the ablation front).}\label{fig:diag}
\end{figure}

\begin{figure}
\includegraphics[width=4cm]{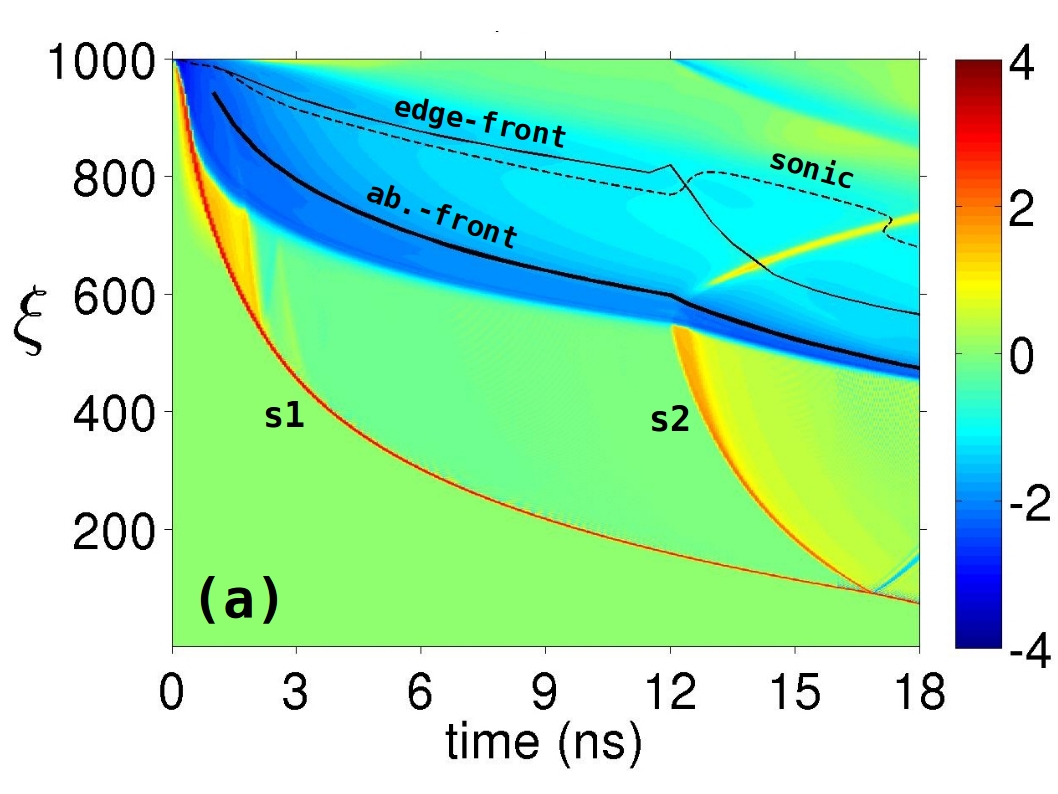}\includegraphics[width=4cm]{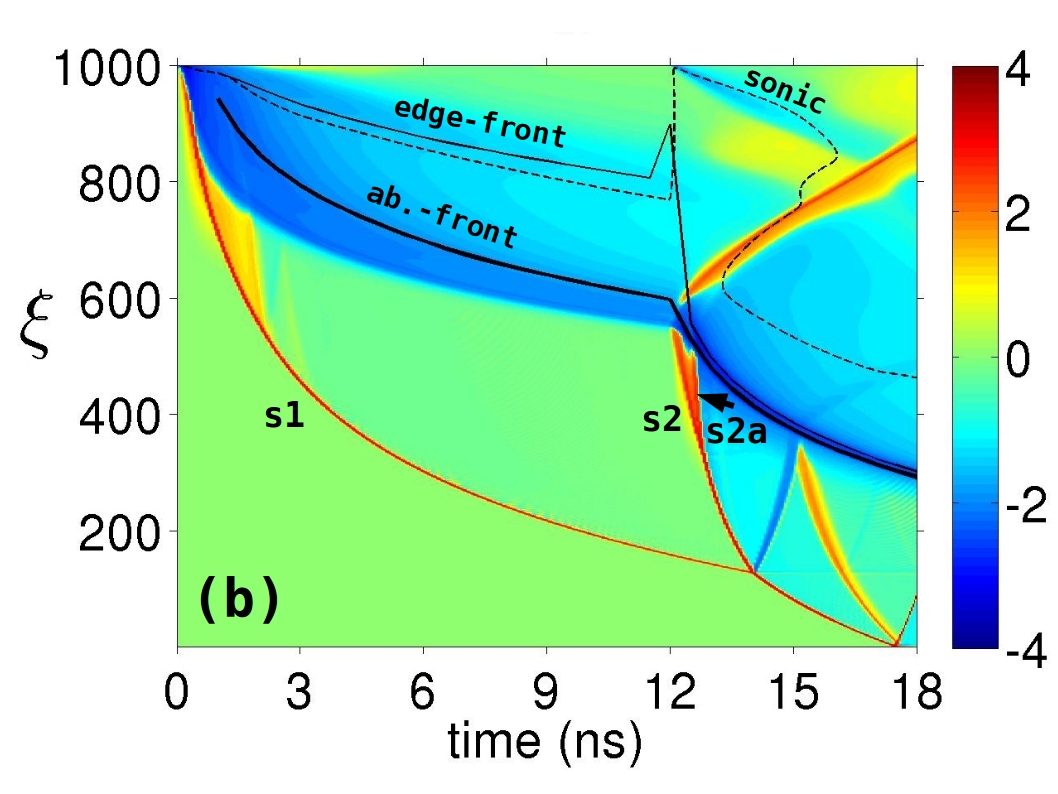}\\
\includegraphics[width=4cm]{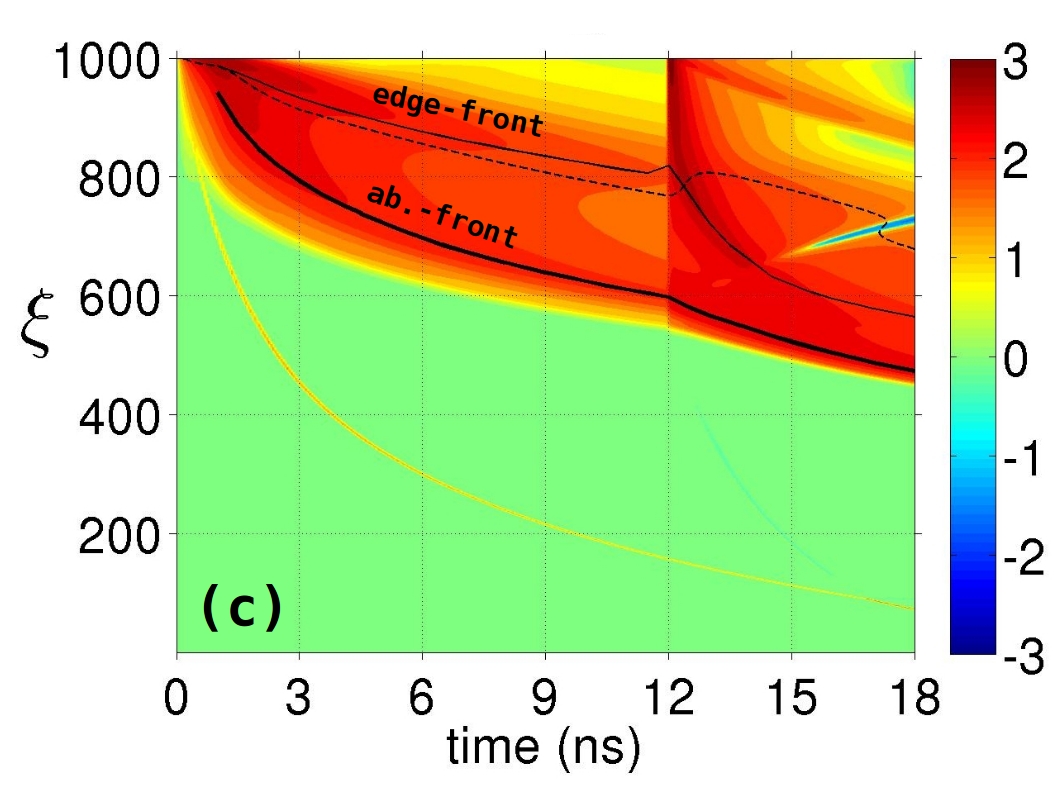}\includegraphics[width=4cm]{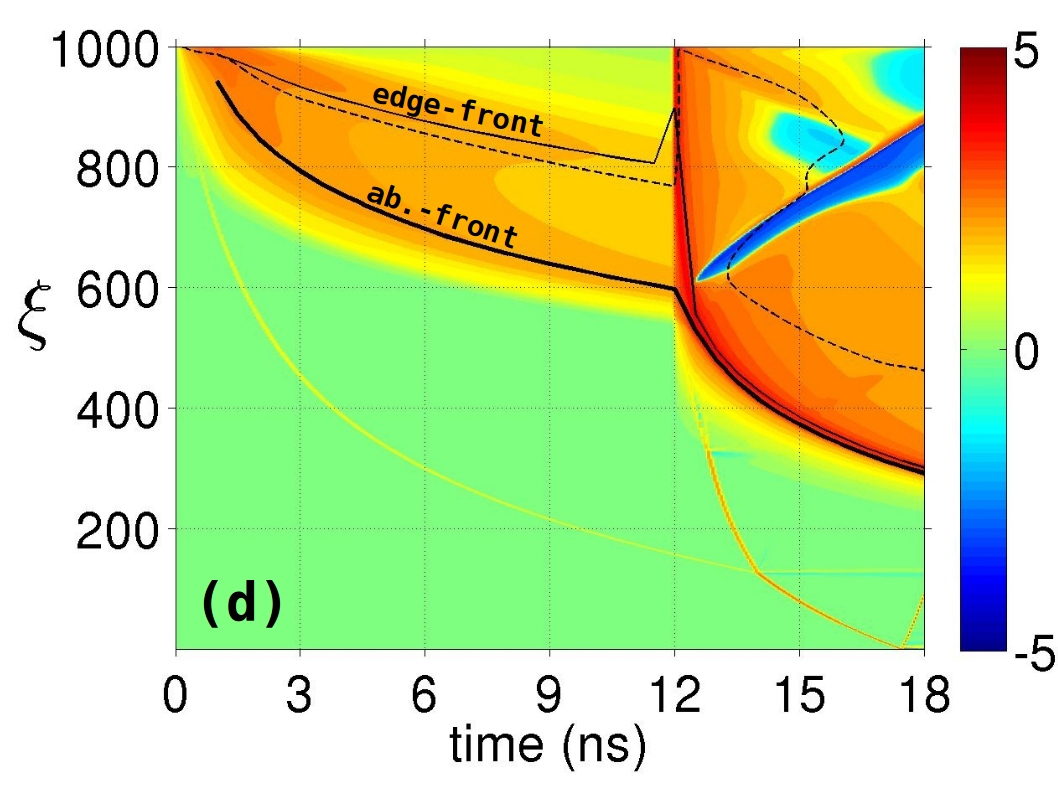}
\caption{(Color online) Same graphs as in Fig.(\ref{fig:diag}) for $T_r^\mathrm{max}$=75 eV (left) and for $T_r^\mathrm{max}$=140 eV (right).}\label{fig:ent}
\end{figure}

\section{Description of the simulated flow} 

The phenomenology of this unwanted shock has been reproduced here in three simple 1D plane simulations with the radiation-hydrodynamic code FCI2 (in multi-group diffusion mode). An incoming black-body ionizing-radiation, with one of the three $T_r$ laws depicted on Fig.(\ref{fig:tr}), impacts a 300 $\mu$m thick plastic ablator ($\rho$=1.05 g/cm$^3$).  Sesame equation of state \cite{eos} and SCO opacity \cite{sco} tables are used. These opacities rely on the self-consistent Dirac average-atom model in the local density approximation and were calculated using the code SCO-RCG. The simulated radiation-spectrum is sampled with 200 equally spaced bins between 0 and 1.5 keV.  

\par

At $t$=0 ns, the x-ray flux hits the outer surface of the ablator (at $\xi$=1000) and builds up an ablation-front (blue area along the thick black solid line). A first shock (s1) is created and propagates through the ablator (from $\xi$=1000 to 100 and below). At $t$=12 ns, the second shock (s2) gets off the ablation front at $\xi$=550 in conjunction with the second $T_r$ rise in Fig.(\ref{fig:tr}). In Kahn's theory, only these two shocks are expected with two $T_r$ rises. 

\par

With a delay (in Fig.(\ref{fig:diag}a)) an unwanted shock (s2a) emerges in the shocked un-ablated material at $t$=14 ns and $\xi$=500 in the case where $T_r^\mathrm{max}$=105 eV. It is important to notice (red arrow in Fig.(\ref{fig:diag}a)) that s2a seems to have been propagating in the ablated material before emerging. In the case where $T_r^\mathrm{max}$=75 eV or 140 eV (Fig.(\ref{fig:ent}a-b)), the fate of the unwanted shock (s2a) is drastically different. It is squarely missing when $T_r^\mathrm{max}$=75 eV and it is strong enough, when $T_r^\mathrm{max}$=140 eV, to outrun shock 2 (s2) giving the false impression that it is missing whereas it is dominant. This odd behavior originates from the existence of a second radiation front in a carbon-based ablator.

\section{A second I-front downstream of the classical ablation front} 

The rate of specific entropy change is represented on Fig.(\ref{fig:diag}b). Although it is mixed with reversible contribution, the strongest positive values are associated mainly with irreversible processes such as heat transfer and ionization as will be demonstrated later on. Clearly, the line called {\it ab.-front} (for classical ablation-front) follows a maximum of entropy production as time goes by. This is consistent with the fact that a significant fraction of the radiation energy carried by the x-rays is absorbed at that location and turns irreversibly into internal energy. Oddly enough, a second line follows a maximum of entropy production (under the name {\it edge-front} in Fig.(\ref{fig:diag}b)). Therefore, a second significant entropy deposition is at stake in the ablated material downstream of the ablation-front. Its reason is to be found in the penetration depth of {\it low} and {\it high} energy photons.

\begin{figure} [t]
\includegraphics[width=7cm]{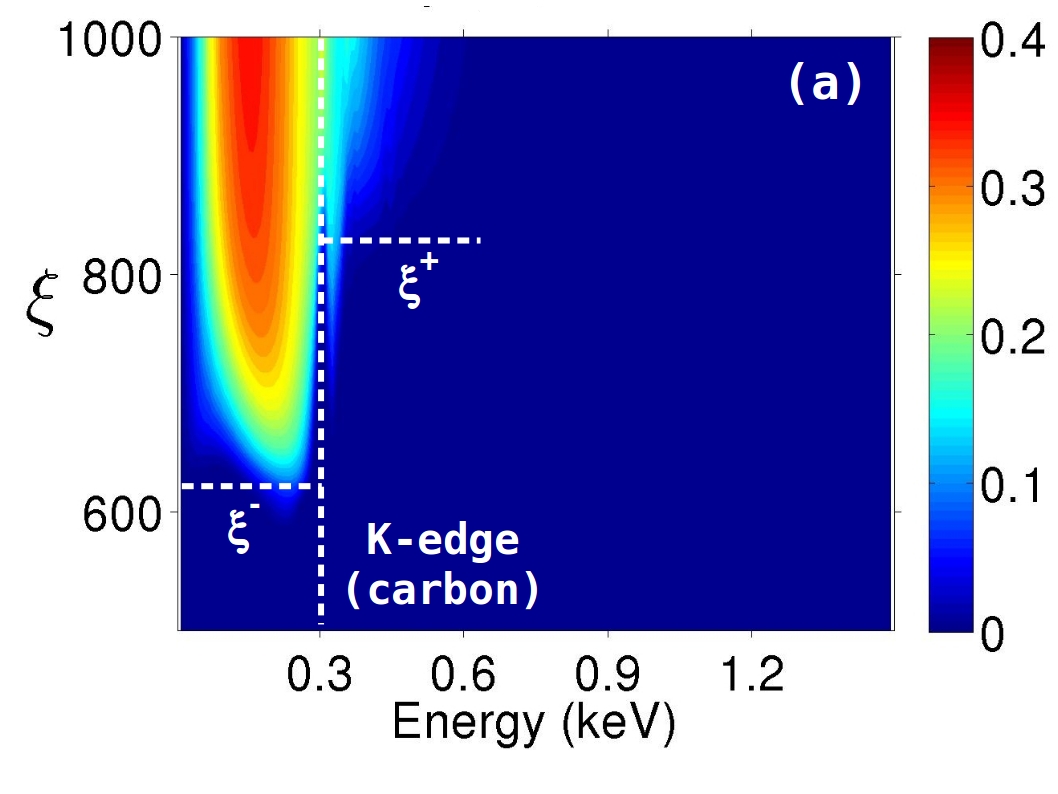}\\
\includegraphics[width=7cm]{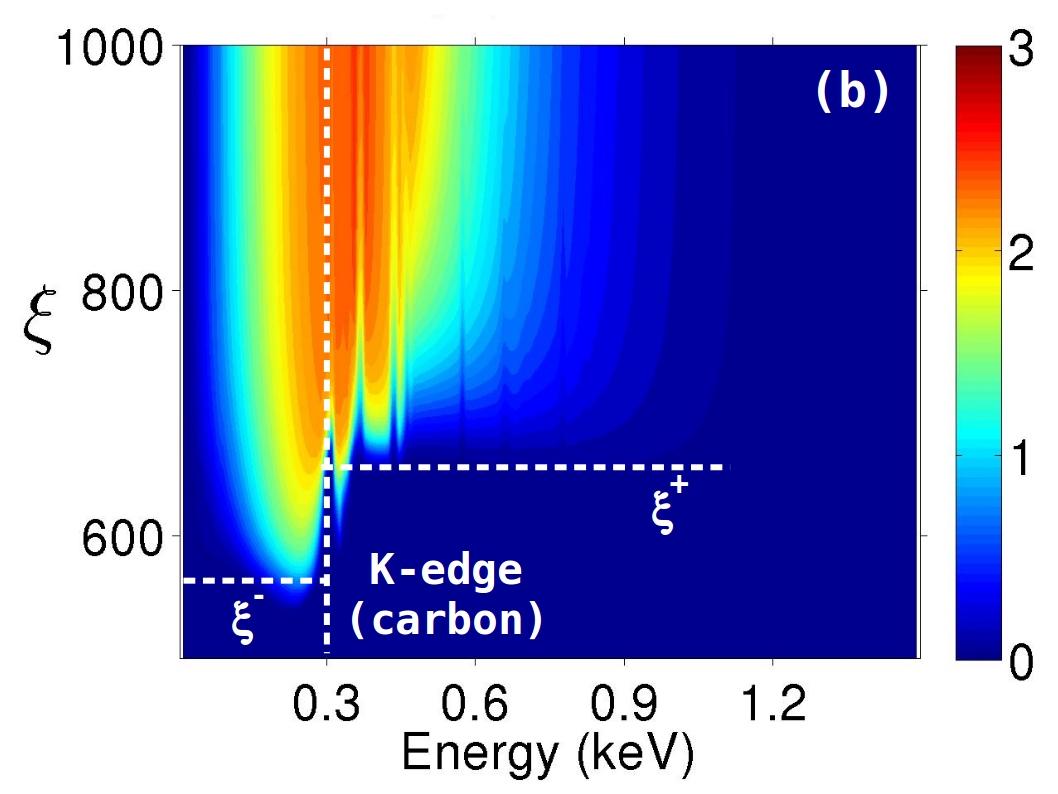}
\caption{(Color online) Spectral intensity (in color) of the radiation coming from the Hohlraum at t=9 ns (a) and t=12.5 ns (b) and going through the ablator with respect to photon energy (horizontal axis) and to lagrangian depth in the ablator (vertical axis). Photons with energy below the K-edge of carbon penetrate deeper in the ablator than photons with energy above.}\label{fig:spec2d}
\end{figure}

\section{Penetration depth of photons versus energy} 

In order to understand how that second front builds up, it is instructive to plot the radiation spectral energy density at two instants (t=9 ns in Fig.(\ref{fig:spec2d}a) and t=12.5 ns in Fig.(\ref{fig:spec2d}b)) with respect to the (lagrangian) depth in the ablator, $\xi$. The black-body radiation that hits the ablator at $\xi$=1000 is clearly split in two in both cases. Photons with energy below 0.3 keV go deeper in the ablator than photons with energy above 0.3 keV. In other words, photons with ``low" energy (slightly less than 0.3 keV) propagate further than ``high" energy photons (slightly more than 0.3 keV). 

\par

\begin{figure}[t]
\includegraphics[width=7cm]{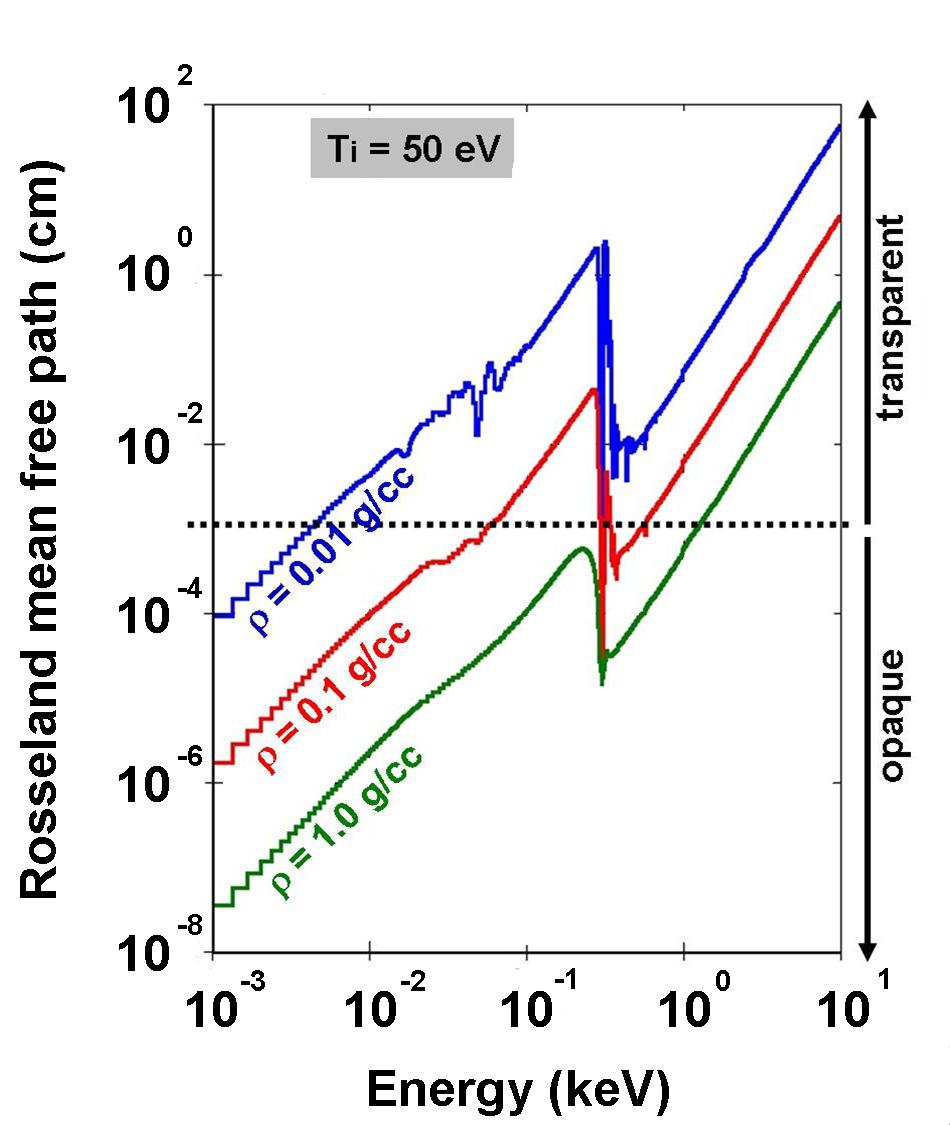}
\caption{(Color online) Evolution of carbon Rosseland mean free path as the density increases for a fixed material temperature of 50 eV. The black dotted line stands for the separation of an ablator that is opaque to the radiation to a transparent ablator.} \label{fig:opac}
\end{figure}

At first glance, this result may seem counter intuitive since photons with high energy (for which matter is transparent) should propagate further than low energy photons (for which matter is opaque). This statement is accurate when opacity varies monotonically (increasing) with respect to energy. This is not the case for plastic whose opacity mainly stems from that of carbon which has a strong absorption K-edge around 0.3 keV (Fig.(\ref{fig:opac})). 

\par

The averaged penetration depths in the ablator of photons with energy E$<$ 0.3 keV and E$>$ 0.3 keV, respectively $\xi^-(t)$ and $\xi^+(t)$, is plotted on Figs.(\ref{fig:diag}) and (\ref{fig:ent}). They are inferred from the energy-dependent depths, such as those in Fig.(\ref{fig:spec2d}a-b), and weighed by the energy density spectrum of the incoming radiation below 0.3 keV (for $\xi^-$) and above 0.3 keV (for $\xi^+$). The thick solid line ({ab.-front}) in Figs.(\ref{fig:diag}) and (\ref{fig:ent}) corresponds in fact to $\xi^-$ whereas the thin solid line ({edge-front}) corresponds to $\xi^+$. As a consequence, the locus of maximum entropy production rate (described earlier) corresponds exactly to the locus of radiation absorption for {\it low} energy (below 0.3 keV) and {\it high} energy photons (above 0.3 keV) as it should. This difference in penetration depth finds an explanation in the inspection of the total spectral mean free path of carbon.

\section{An absorption edge around the maximum of the x-ray spectrum causes the formation of two radiation fronts} 

In Fig.(\ref{fig:opac}), the spectral mean free path is plotted at different density for a temperature of matter $T=50$ eV which is typical, in an actual ICF drive, in the ablated region at the second $T_r$ rise. 

\par

A fiducial horizontal dotted line has been drawn around a Rosseland mean free path of tens of microns ($0.001$ cm) to separate between transparent and opaque behaviors in an ablator of width of order hundreds of microns. As a result, the low density ablated CH with $\rho\sim0.01$ g/cm$^3$ at $\xi$=1000 is transparent to incoming photons with energy above a few eV (blue curve on Fig.\ref{fig:opac}). As photons go deeper towards the main ablation front, they go through denser and denser ablated material up to few g/cm$^3$ in the neighborhood of the ablation front. On one hand, photons of energy just above 0.3 keV will propagate {\it freely} in the ablator until they find matter at $\rho\approx 0.1$ g/cm$^3$ on their way (red curve on Fig.\ref{fig:opac}). On the other hand, photons of energy just below 0.3 keV will propagate {\it freely} until they reach matter at $\rho\approx 1$ g/cm$^3$ (green curve on Fig.\ref{fig:opac}). This means that, photons with energy just below 0.3 keV will be stopped at a location where $\rho\approx 1$ g/cm$^3$ (it corresponds to the location of the main ablation front at $\xi^-(t)$) and photons with energy just above 0.3 keV will be stopped in the ablated region at a location where $\rho\approx 0.1$ g/cm$^3$ (located downstream from the ablation front at $\xi^+(t)$). Therefore, the K-edge of carbon is responsible for two distinct regions where radiation is absorbed in the ablator: one corresponding to the classical ablation front ($\xi^-$) and another one ($\xi^+$), in the ablated material, that we call an {\it edge-front} for obvious reasons now. 

\par

\begin{figure}[t]
\includegraphics[width=7cm]{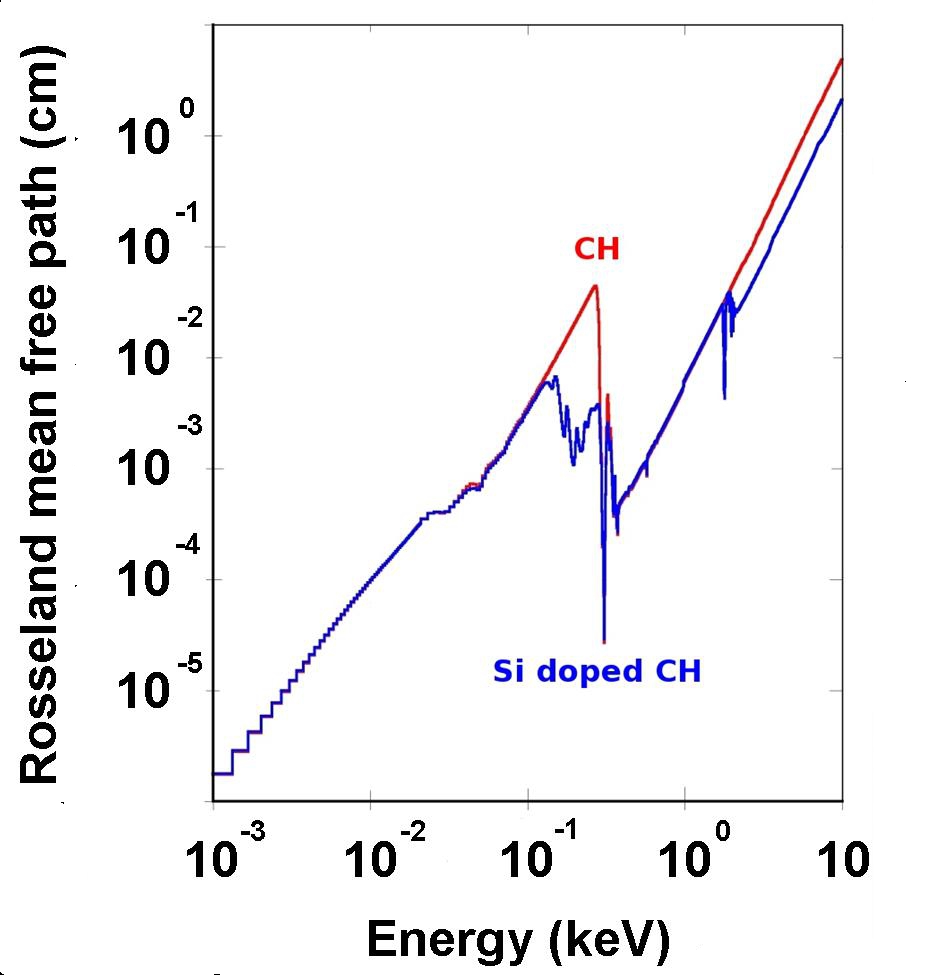}
\caption{(Color online) Carbon Rosseland mean free path compared to Si doped CH Rosseland mean free path at a given material temperature of 50 eV and a density $\rho=0.1$ g/cm$^3$. The K-edge of pure carbon is almost washed out by the presence of Si dopent.}\label{fig:opacsi} 
\end{figure}

The position of $\xi^+$ in the ablated material is tied to the position of $\rho\approx 0.1$ g/cm$^3$. That position, in turn, depends on the mass ablation rate and on the underdense eos of the ablator. The formation of this edge-front is not an artifact of our simulation code FCI2. We have performed these same simulations with Si doped CH, instead of pure CH, for which the K-edge is almost washed out in the Rosseland mean free path (Fig.(\ref{fig:opacsi})). The resulting {\it edge-front} was barely visible and the extra shock 2a vanished irrespective of the choice of the $T_r$ law. 

\section{The edge-shock} 

Khan's theory can be applied separately to the two fronts corresponding to photons with $E>$0.3 keV and $E<$0.3 keV. After the second $T_r$ rise, each front generates one shock of its own. The ablation front generates (s2) whereas the edge-front in the ablated region generates the unwanted shock (s2a). This is the reason why we have called it an {\it edge-shock} throughout this article. It can catch up with the classical ablation front after the second $T_r$ rise because the thermodynamic conditions in the ablated region become favorable: before the second $T_r$ rise  the sonic point relative to the ablation front does not extend far enough (thick dashed line in Fig.(\ref{fig:diag}a-b)) in the ablated region for the edge-front to have any causal relation with the unablated region. After the second $T_r$ rise, on the other hand, the sonic point is moved further behind the edge-front. Accordingly, a perturbation generated by the edge-front can propagate upstream, catch up with the ablation front and emerge in the unionized shocked region and eventually catch up with s2 (before or after s2 intersects with the trajectory of s1). Therefore, the creation of an edge-shock can be put down to the existence of an edge-front (if the spectrum of the ionizing-radiation overlaps with an absorption edge of the ablator opacity) and to the thermodynamic state of matter in the ablated region (the edge-front must be in the sonic region of the ablation front). \par 
As ill luck would have it, in the ICF context, the K-edge of carbon ($E_K=$285 eV exactly) lies at the center of the broad spectrum of x-rays coming from the hohlraum submitted to a radiative temperature $T_r\approx$100 eV that is typical around the second rise of the radiative temperature history in an actual Holhraum (the maximum of the x-ray distribution is roughly at $E_\mathrm{max}=$2.8$\times T_r=$280 eV$\approx E_K$). This is the reason why the edge-shock (s2a) is well separated from the ablation-shock (s2) in Fig.(\ref{fig:diag}) at a final $T_r=$105 eV whereas it does not seem to appear at higher ($T_r$=130 eV) or lower ($T_r=$80 eV) radiative temperature. At lower $T_r$, the majority of the ionizing-radiation is well below $E_K=$0.3 keV : the classical ablation-front is much stronger than the edge-front when it comes to generating a shock and the flow eventually looks like there is only one single shock (s2$\gg$s2a). It is the other way around for higher $T_r$ where the majority of the ionizing-radiation is well above $E_K=$0.3 keV. The flow still looks like there is only one single shock (s2$\ll$s2a) but, this time, the edge-front is much stronger than the classical ablation front. If the $T_r$ was increased yet further, the edge-front would turn continuously into a classical ablation front as the spectrum of x-rays would shift to higher energies away from the absorption edge of the ablator.\par 
This effect should certainly be looked upon in the astrophysical context in situations where a flux of photons with radiative temperature $T_r$ goes through an ISM made of matter with a strong absorption-edge at energy $E_{\mathrm{edge}}$. If the flux is intense enough and if $T_r\approx E_{\mathrm{edge}}/$2.8, an edge-front will form along with the ablation-front. As an example, the formation of O-star involves hydrogen ($E_{\mathrm{edge}}=$13.6 eV). An ionizing-radiation at $T_r$=13.6/2.8=4.9 eV=56000K could  be responsible for the formation of two shocks in the hydrogen cloud. Such a radiative temperature corresponds to the particular class of O7 stars.

\end{document}